\documentclass[final,3p,times]{elsarticle}

\usepackage{amssymb}

\usepackage{graphicx}

\journal{Elsevier}

\begin{document}
\begin{frontmatter}
\title{Critical exponents from large mass expansion
}
\author{Hirofumi Yamada}
\ead{yamada.hirofumi@it-chiba.ac.jp}

\address{
Division of mathematics and science, Chiba Institute of Technology, 
\\Shibazono 2-1-1, Narashino, Chiba 275-0023, Japan} 
\begin{abstract}
We perform estimation of critical exponents via large mass expansion under crucial help of $\delta$-expansion.   We address to the three dimensional Ising model at high temperature and estimate $\omega$, the correction-to-scaling exponent, $\nu$, $\eta$ and $\gamma$ in unbiased and self-contained manner.  The results read at the highest $25$th order expansion $\omega=0.8002$, $\nu=0.6295$, $\eta=0.0369$ and $\gamma=1.2357$.  Estimation biased by $\omega=0.84(4)$ is also performed and proved to be in agreement with the summary of recent literatures.
\end{abstract}
\date{\today}

\end{frontmatter}

  
\section{Introduction}
It would be a common understanding that the scaling corrections in three dimensional Ising models play an important role in the estimation works of critical exponents as noted by Weger \cite{weg}.   Since then people published the results on the estimation of the correction-to-scaling exponent and made use of it in the computation of various leading exponents.   For many contributions to this subject, see for example ref.\cite{peli} and recent literatures \cite{ari,den,ber,pog,has,lit,gor,she}.  

To the best of our knowledge, most of the unbiased computation has been carried out in field theoretic models and computer simulations.  For instance in Ising universality class at three dimension, there exist only a few literatures \cite{gut,butera1} where pure series expansions (high and low temperature expansions without universality hypothesis) of Ising models themselves were employed for the full exponents computation in a self-contained manner.  Rather, one performs the estimation of critical quantities under the bias of the inverse critical temperature $\beta_{c}$ and/or the correction-to-scaling exponent $\theta$.   This trend may be due to the lack of an unbiased accurate computational framework in the high and low temperature expansions.   To cover the deficiency, we like to propose an unbiased and self-contained approach to critical exponents in the frame work of series expansions.  
As a crucial test, we attempt estimations of $\omega=\theta/\nu$, $\eta$, $\nu$ and $\gamma$ in the cubical Ising model.

The Ising model we consider is specified by the Boltzman weight $\exp(\beta\sum_{<i,j>}s_{i}s_{j})$, $s_{i}^2=1$,  
where $\beta$ denotes the inverse temperature and $<i,j>$ means the pair of sites $i$ and $j$ which are connected by single link.   In literatures, the critical behaviors of thermodynamic quantities is discussed in terms of the reduced temperature 
$\tau=1-\beta/\beta_{c}$ \cite{camp}.  
However we adopt an appropriate mass square $M$ to describe the thermodynamic quantities.  The description due to $\tau$ needs $\beta_{c}$, but $M$ has the trivial critical point $M=0$.  Hence, computing critical exponents results from the asymptotic behavior of quantities of interest in the massless limit, and importantly, the job is bias-free from the determination of $\beta_{c}$.   For instance, we note that even the critical temperature is given by the limit 
$\beta_{c}=\lim_{M\to 0}\beta(M)$.  Actually, 
the explicit estimations of $\beta_{c}$ and critical exponents $\nu$ and $\gamma$ in the square Ising model was performed in ref.\cite{yam3} based upon the $1/M$ expansion.  Also in the 3-dimensional model, ref.\cite{yam} attempted to compute critical temperature and exponents under the preliminary version of the method to be presented in this paper.  We follow the approach with refinements and demonstrate that the accuracy of estimate in the new protocol is highly improved.  One point of the refinements consists of the re-formulation of parametric extension of thermodynamic quantities \cite{yam} by the linear differential equation (LDE) with constant coefficients \cite{yam3}.  Next point is to access directly to the critical exponents by the consideration of the ratio of the first and second derivatives of $\beta(M)$ and the ratio of $\chi(M)$ and its first derivative.  The last point is the use of characteristic structure of Wegner expansion and LDE, leading bootstrapping point of view.  Also in the present analysis, the $\delta$-expansion plays crucial and inevitable roles \cite{yam2,yam3}.

This paper is organized as follows:  In the next section, we investigate structures of series expansions near the critical point for $\beta(M)$, $\chi(M)$ and related functions which shall be  used for the estimation of critical quantities.  Then, in section 3, we present estimation work on the critical exponents in the self-contained and unbiased manner based upon the large mass expansion.  In section 4, an biased estimation will be presented under the use of $\omega=0.84(4)$ \cite{peli}.  Finally we give concluding remarks on this work.

\section{Preliminary studies}
In reference to \cite{peli}, we start the arguments by the consideration of the series expansion of the correlation length $\xi$ at high temperature.  Let the spectrum of powers of corrections be generated by the basic ones, $1$, $\theta$, $\theta_{1}$, $\theta_{2}$, $\cdots$ and their integer multiples.  Then, 
\begin{eqnarray}
\xi&\sim& f\tau^{-\nu}\bigg[1+\Big\{a\tau^{\theta}(1+a_{11}\tau+a_{12}\tau^2+\cdots)+a_{2}\tau^{2\theta}(1+a_{21}\tau+a_{22}\tau^2+\cdots)+\cdots\Big\}\\
& &+\Big\{b\tau^{\theta_{1}}(1+b_{11}\tau+b_{12}\tau^2+\cdots)+b_{2}\tau^{2\theta_{1}}(1+b_{21}\tau+b_{22}\tau^2+\cdots)+\cdots\Big\}+\cdots\cdots+\Big\{u\tau(1+u_{1}\tau+u_{2}\tau^2+\cdots)\Big\}\bigg]\nonumber\\
& &+\xi_{R},\nonumber
\label{xi}
\end{eqnarray}
where $\xi_{R}=const\times \tau(1+r_{1}\tau+r_{2}\tau^2+\cdots)$ stands for the analytic back ground  and $0<\theta<\theta_{1}<\theta_{2}<\cdots$.   The inversion yields series in $\xi/f$ with the spectrum of exponents of the form scaled by $\nu$, $-1/\nu-(m+n\theta+n_{1}\theta_{1}+n_{2}\theta_{2}+\cdots)/\nu$ where $m, n, n_{k}\, (k=1,2,3,\cdots)$ are all non-negative integers.  Leading term is $(\xi/f)^{-1/\nu}$ provided $\theta<1$ and the next term is $(\xi/f)^{-(1+\theta)/\nu}$.  Here, we pose the assumption that $\theta_{1}>1$ and then the third term proves to be $(\xi/f)^{-(1+1)/\nu}=(\xi/f)^{-2/\nu}$.  Then, inversion of the above series to a few orders reads
\begin{equation}
\tau=(\xi/f)^{-1/\nu}+\frac{a}{\nu}(\xi/f)^{-(1+\theta)/\nu}+\frac{b}{\nu}(\xi/f)^{-2/\nu}+\cdots.
\end{equation}
Using $\xi^{-2}\sim M(1+const \times M+\cdots)$ near the critical point, we have $\xi^{-1/\nu}\sim M^{1/2\nu}(1+const\times M+\cdots)$ and the term of order $M^{1/2\nu+1}$ appears.  This term belongs to rather higher orders provided $\nu>1/2$ and can be omitted.   Thus, we arrive at
\begin{equation}
\tau=f^{\frac{1}{\nu}}M^{\frac{1}{2\nu}}(1+\frac{a}{\nu}f^{\frac{\theta}{\nu}}M^{\frac{\theta}{2\nu}}+\frac{b}{\nu}f^{\frac{1}{\nu}}M^{\frac{1}{2\nu}}+\cdots),
\label{tauscale}
\end{equation}
and 
\begin{equation}
\beta=\beta_{c}(1-\tau).
\label{betascaling}
\end{equation}

From here on we denote thermodynamic quantities near the critical point with the lower index $<$ and at high temperature similarly with $>$.   Based upon the large mass expansion of $\beta$ denoted by $\beta_{>}$, we later carry out computation of the critical exponent $\nu$.  For the estimation of $\nu$, the ratio $\beta^{(2)}/\beta^{(1)}:=f_{\beta}$ is convenient since $\nu$ itself appears as the leading term,
\begin{equation}
\frac{\beta_{<}^{(2)}}{\beta_{<}^{(1)}}=f_{\beta<}=-\frac{1}{2\nu}-A_{1} x^{-\frac{\theta}{2\nu}}-A_{2}x^{-\frac{1}{2\nu}}+\cdots,
\label{ratio3d}
\end{equation}
where
\begin{equation}
x:=1/M,
\end{equation}
and
\begin{equation}
\beta^{(\ell)}=\Big(\frac{d}{d\log x}\Big)^{\ell}\beta.
\end{equation}
The amplitude is written by $a$, $\nu$, $\theta$ and so on but its detail is not relevant for our purpose.   

The magnetic susceptibility $\chi$ is defined by
\begin{equation}
\chi=\sum_{n:sites}< s_{0}s_{n}>.
\end{equation}
The re-writing of $\chi$ in terms of $x$ instead of $\tau$ is straightforward:  In the critical region, it suffices to substitute $\tau(x)$ in (\ref{tauscale}) into the standard expression of $\chi_{<}$ like (\ref{xi}), $\chi_{<}(\tau)\sim C\tau^{-\gamma}[1+const(\tau^{\theta}+\cdots)+\cdots+const(\tau+\cdots)]+\chi_{R}$ ($\chi_{R}$ denotes the analytic back ground) \cite{aha}.  Substituting (\ref{tauscale}) and further recasting the series in $x$, one obtains
\begin{equation}
\chi_{<}=Cx^{\frac{\gamma}{2\nu}}(1+const\cdot x^{-\frac{\theta}{2\nu}}+const\cdot x^{-\frac{1}{2\nu}}+\cdots).
\end{equation}
Scaling relation due to Fisher \cite{fish} tells us that $\gamma/(2\nu)=1-\eta/2$ and what we can directly measure is $\eta$ rather than $\gamma$.   In the estimation of $\eta$, it proves convenient to address $(\log\chi)^{(1)}=(d/d\log x)\chi:=f_{\chi}$.   It behaves near the critical point,
\begin{equation}
f_{\chi<}=\frac{\gamma}{2\nu}+const\cdot x^{-\theta/2\nu}+const\cdot x^{-1/2\nu}+\cdots.
\label{chiscaling}
\end{equation}
All behaviors of the series (\ref{betascaling}), (\ref{ratio3d}) and (\ref{chiscaling}) can be written as
\begin{equation}
f_{<}=f_{0}+c_{1}x^{-\lambda_{1}}+c_{2}x^{-\lambda_{2}}+\cdots.
\label{general_scaling}
\end{equation}
This satisfies linear differential equation (LDE) to $K$th order
\begin{equation}
\prod_{n=0}^{K}\Big[\lambda_{n}+\frac{d}{d\log x}\Big]f_{<}=O(x^{-\lambda_{K+1}}), \quad \lambda_{0}=0.
\label{lde}
\end{equation}
Integration over $\log x$ corresponding to the operator $\lambda_{0}+d/d\log x=d/d\log x$ gives
\begin{equation}
\prod_{n=1}^{K}\Big[1+(\lambda_{n})^{-1}\frac{d}{d\log x}\Big]f_{<}=f_{0}+O(x^{-\lambda_{K+1}}).
\label{basiclde}
\end{equation}
The spectrum of exponents are thus interpreted as the spectrum of the roots of the characteristic equation associated with the LDE.  The non-universal amplitudes (and also the inverse critical temperature $\beta_{c}$) appear as the integration constants. 

In our approach, we use (\ref{basiclde}) to extract $f_{0}$ and exponent $\lambda_{n}$ for low $n$.  Here, the point is that the series $f_{<}$ valid in the critical region is not explicitly known.  It would be nice if we could use $f_{>}$ in the place of $f_{<}$ but there is no matching region of respective expansions.   It is a crucial observation, therefore, that $\delta$-expansion \cite{yam2,yam,yam3} removes the obstruction by the dilatation of the scaling region.  Suppose one has expansion truncated at order $N$, $f_{N>}(M)=\sum_{n=0}^{N}a_{n}(1/M)^n$.  The dilatation can be installed by the change of the argument from $x$ to $t$ defined by $x=t/(1-\delta)$ $(0\leq \delta \leq 1)$ and the expansion to an appropriate order depending on the truncation order $N$.   
After the limit $\delta\to 1$, which stands formally for the infinite magnification, these operations give the transformation
\begin{equation}
x^{n}\to C_{N, n}t^n,\quad (n=0,1,2,3,\cdots,N)
\end{equation}
where $C_{N,n}$ stands for the binomial coefficient $C_{N,n}=N!/\{n!(N-n)!\}$.  
Denoting the transform due to the $\delta$ expansion by $D_{N}$, one thus has $D_{N}[x^n]=C_{N,n}t^n$ and
\begin{equation}
D_{N}[f_{N>}]=\sum_{n=0}^{N}a_{n}C_{N,n} t^n:=\bar f_{N>}.
\label{f_>}
\end{equation}
Actually, in some models (see \cite{yam2,yam3}), the resulting series in $t$ have been confirmed to exhibit known scalings.  Also in the cubic Ising model, we observe the scaling behaviors of $D_{N}[f_{\beta>}]=\bar f_{\beta N>}$ and $D_{N}[f_{\chi>}]=\bar f_{\beta N>}$ as shown in Fig. 1.  In addition, their derivatives which tend to zero exhibit the behaviors.  For example, $D_{N}[f_{\beta N>}^{(\ell)}]$ for $\ell=1,2,3$ shows the scaling to zero in the region $\sim (0.10,0.12)$ and also $D_{N}[f_{\chi N>}^{(\ell)}]$ for $\ell=1,2,3,4$ in $\sim (0.11,0.12)$.  Though the region of scaling is narrow, we expect matching in these regions of $D_{N}[f_{\beta(\chi) N>}^{(\ell)}]$ with the transformed $f_{\beta(\chi)<}^{(\ell)}$.

The matching requires the transformation of $f_{<}(M)$.  Though clear basis of $\delta$ expansion for $f_{<}$ valid in the critical region is not obtained yet, we proceed by assuming the term $x^{-\lambda}$ for positive real $\lambda$ appearing the expansion in the critical region (see (\ref{general_scaling})) also receives $N$ dependent coefficient
\begin{equation}
C_{N,-\lambda}=\frac{\Gamma(N+1)}{\Gamma(-\lambda+1)\Gamma(N+\lambda+1)},
\end{equation}
which is just the analytic extension of $C_{N,n}$.  This tells us that $D_{N}[x^{-\lambda}]=0$ for positive integer $\lambda$ and this has been confirmed in the square Ising model \cite{yam3}.  Thus, we have from (\ref{general_scaling})
\begin{equation}
D_{N}[f_{<}]=\bar f_{N<}=f_{0}+c_{1}C_{N,-\lambda_{1}}t^{-\lambda^{'}_{1}}+c_{2}C_{N,-\lambda_{2}}t^{-\lambda^{'}_{2}}+\cdots,
\end{equation}
and all $\lambda^{'}_{i}$ $(i=1,2,3,\cdots)$ can be understood as non-integers (We should consider that $\{\lambda_{n}^{'}\}$ is a subset of $\{\lambda\}$ from which positive integer exponents are removed away).  A benefit of transformed one is found in the large order behavior of $C_{N,-\lambda}$.  As $N\to \infty$, one finds 
$C_{N,-\lambda}\to N^{-\lambda}/\Gamma(-\lambda+1)$ 
and, for positive $\lambda$ appearing in the critical region, the created amplitude $C_{N,-\lambda}$ decreases toward zero.  This gives rise the effect that the corrections to the asymptotic scaling, for example appeared in (\ref{ratio3d}) and (\ref{chiscaling}), decrease with the order $N$ at fixed finite $t$ such that $(t N)^{-\lambda^{'}_{i}}\to 0$.  

One might feel the lack of firm and rigorous basis on the $\delta$ expansion of the series in the critical region.  However, we emphasize that what really affects the estimation task is that the leading constant $f_{0}$ and the critical exponents $\lambda^{'}_{i}$ for $i=1,2,3,\cdots$ are left invariant under the $\delta$ expansion.  The detailed information of the modified amplitude does not affect our results since the LDE to be satisfied does not concern with the amplitudes.

Now, LDE to be satisfied by $\bar f_{N<}$ thus becomes
\begin{equation}
\Big(\prod_{n=0}^{K}L_{n}(t)\Big)\bar f_{N<}=O(t^{-\lambda^{'}_{K+1}}),
\label{deltalde}
\end{equation}
where
\begin{equation}
L_{n}(t)=\lambda^{'}_{n}+\frac{d}{d\log t},\quad \lambda^{'}_{0}=0.
\end{equation}
Our estimation protocol is based upon (\ref{deltalde}).  In the place of $\bar f_{N<}$, $\bar f_{N>}$ (see (\ref{f_>})) will be substituted because the scaling behavior is observed in it, and $f_{0}$ and unknown critical exponent $\lambda^{'}_{i}$ will be estimated.    
\begin{figure}
\centering
\includegraphics[scale=0.8]{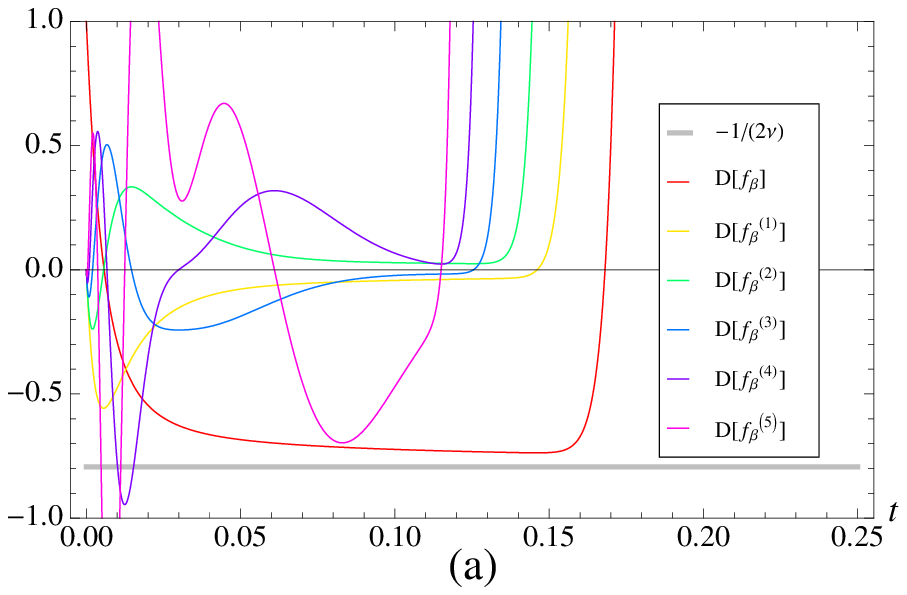}
\includegraphics[scale=0.8]{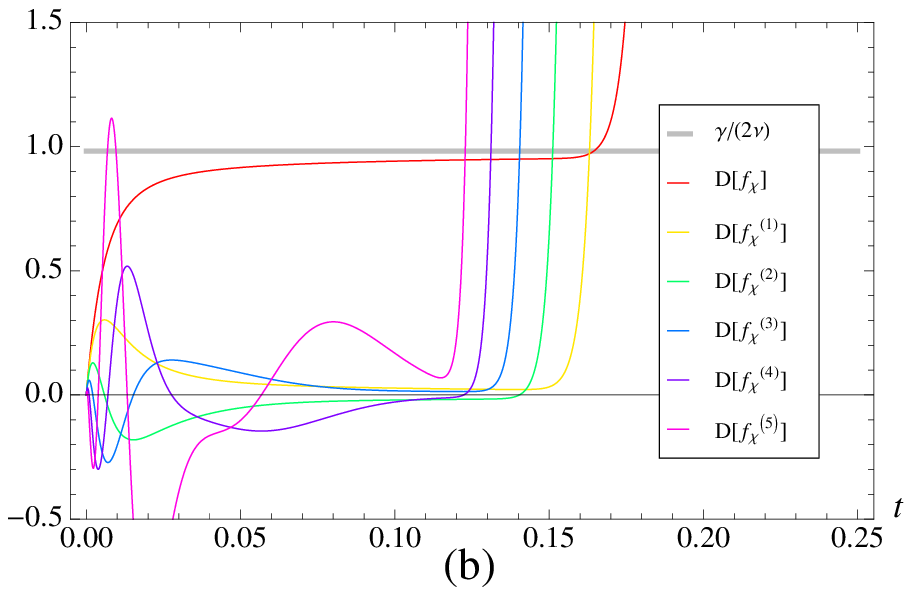}
\caption{The graphs (a) and (b) show plots of $\bar f_{\beta>}$ and $\bar f_{\chi>}$ at respective highest orders (N=24,25) and their derivatives with respect to $d/d\log t$ to the 5th order.  The gray lines indicate $-1/(2\nu)$ for $\nu=0.6301$ in (a) and $\gamma(2\nu)$ for $\gamma=1.2373$ and $\nu=0.6301$ in (b).}
\end{figure}

\section{Unbiased estimation}
Now we turn to the estimation of critical exponents by making use of the large mass expansion.  We adopt the second moment mass $3\chi/\mu$ as the basic parameter $M$, $M=3\chi/\mu$.  
As for the second moment mass, its high temperature expansion in $\beta$ can be calculated by using the result of Butera and Comi for the magnetic susceptibility $\chi$ and the second moment $\mu$ up to $25$th order \cite{butera}.  Combining the results and inverting $M(\beta)$, we have 
\begin{eqnarray}
\beta&=&x-6x^2+\frac{124}{3}x^3-312x^4+\frac{12596}{5}x^5-21432 x^6+\frac{1330848}{7}x^7-1745344 x^8+\frac{148384348}{9}x^9\nonumber\\
& &-\frac{797787336}{5}x^{10}+\frac{17341288504}{11}x^{11}-15857888272 x^{12}+\frac{2106367479672}{13}x^{13}-\frac{11748802870160}{7}x^{14}\nonumber\\
& &+\frac{263968267347944}{15}x^{15}-186504592354608 x^{16}+\frac{33924951987330804}{17}x^{17}-21535692193295224 x^{18}\nonumber\\
& &+\frac{4449606807205690200}{19}x^{19}-\frac{12821205881021198992}{5}x^{20}+\frac{197756701920466780928}{7}x^{21}\nonumber\\
& &-\frac{3442869826889278353376}{11}x^{22}+\frac{80156432259652309452520}{23}x^{23}-\frac{116948936021276297965072}{3}x^{24}\nonumber\\
& &+\frac{10946582972904015563857296}{25}x^{25}+O(x^{26}).
\label{beta_M}
\end{eqnarray}
Then from $f_{\beta>}=\beta^{(2)}_{>}/\beta^{(1)}_{>}$, 
\begin{equation}
f_{\beta>}=1-12 x+104 x^2-1008 x^3+10416 x^4-\cdots.
\label{beta_high}
\end{equation}
and the $\delta$-expansion at the expansion order $N$ transforms the above result into
\begin{equation}
D_{N}[f_{\beta N>}]=\bar f_{\beta N>}=1-12 C_{N,1}t+104C_{N,2} t^2-1008 C_{N,3} t^3+\cdots,
\label{beta_high}
\end{equation}
where the last term should be of the order $t^N$.  We note that the highest order of $f_{\beta>}$ is $24$th which comes from $25$th order $\beta_{>}$.

The susceptibility is also written in $x$, by the substitution of $\beta_{>}(x)$ into $\chi_{>}(\beta)$ obtained in \cite{butera}, giving
\begin{eqnarray}
\chi_{>}&=&1+6x-6x^2+36x^3-270x^4+2268x^5-20436x^6+193176x^7-1890462x^8+18990892x^9-194709708x^{10}\nonumber\\
& &+2029271688x^{11}-21435300372x^{12}+228983179752x^{13}-2469626018184x^{14}
+26855777435248x^{15}\nonumber\\
& &-294145354348974x^{16}+3242105906258220x^{17}-35935261094616124x^{18}
+400295059578038760x^{19}\nonumber\\
& &-4479014443566807276x^{20}+50319506857313420376x^{21}-567383767790459777016x^{22}\nonumber\\
& &+6418899321986117552400x^{23}-72838651914163555355012x^{24}+828839976149614386374184x^{25}-\cdots,
\end{eqnarray}
and from $f_{\chi>}=\chi^{(1)}_{>}/\chi_{>}$,
\begin{equation}
f_{\chi>}=6x-48x^2+432x^3-4167 x^4+42336 x^5-\cdots.
\label{chi_high}
\end{equation}
We thus arrive at
\begin{equation}
D_{N}[f_{\chi N>}]=\bar f_{\chi N>}=6 C_{N,1}x-48 C_{N,2}x^2+432 C_{N,3}x^3-4167 C_{N,4} x^4+\cdots.
\label{chi_high}
\end{equation}

We denote the exponents appearing in (\ref{ratio3d}) and (\ref{chiscaling}) be $p_{0}=0, p_{1}, p_{2}, p_{3}, \cdots$.  Further, let us omit the subscript $N$ in $\bar f_{\beta(\chi) N>(<)}$ for the sake of notational simplicity.   In both expansions of $\bar f_{\beta <}$ and $\bar f_{\chi <}$, the leading correction is given by $\sim x^{-p_{1}}=x^{-\theta/2\nu}=x^{-\omega/2}$.  So we first address to the estimation of $\omega$.  Although both of $\bar f_{\beta>}$ and $\bar f_{\chi>}$ seem to be useful, the function $\bar f_{\chi>}$ has an advantage that the behaviors of derivatives show clearer expected scalings than the derivatives of $\bar f_{\beta>}$ (see FIG. 1).   For instance, $\bar f_{\chi>}^{(4)}$ exhibits rough scaling but $\bar f_{\beta>}^{(4)}$ does not.  It is therefore appropriate to make use of $\bar f_{\chi>}^{(k)}$ $(k=0,1,2,3, 4)$.   The order $K$ of LDE is actually limited according to the order of expansion.  From FIG.1(b), we observe that $\bar f_{\chi>}$-derivatives to 4th order show rather reliable scalings, whereas 5th order does not.  The recipe in our protocol depends further on the number of unknown parameters ($\gamma/2\nu$, $p_{1}$, $p_{2}$, $\cdots$) involved.  Less number of parameters needs less order of the derivatives.  Then the most appropriate choice is found to be  $K=2$ LDE ($2$ndLDE) with the incorporation of two correction terms $x^{-p_{1}}$ and $x^{-p_{2}}$.  Neglecting higher order corrections, we employ the ansatz $f_{\chi<}=\gamma/(2\nu)+const\cdot x^{-p_{1}}+const\cdot x^{-p_{2}}$ and then consider the second order version of (\ref{deltalde}),
\begin{equation}
\hat L_{1}\hat L_{2}\bar f_{\chi<}=\frac{\gamma}{2\nu},
\label{lde_chi}
\end{equation}
where $p_{1}=\omega/2$, $p_{2}=1/2\nu$ and
\begin{equation}
\hat L_{n}=1+p_{n}^{-1}\frac{d}{d\log t}.
\end{equation}
In the vicinity of $t$ at which the above LDE holds locally, the expansion of $\bar f_{\chi<}$ is given by shifting $\log t\to \log t+\epsilon$ as
\begin{equation}
\hat L_{1}\hat L_{2}\bar f_{\chi<}(t e^{\epsilon})
=\hat L_{1}\hat L_{2}\bar f_{\chi<}(t)+\hat L_{1}\hat L_{2}\bar f^{(1)}_{\chi<}(t)\epsilon+
 \hat L_{1}\hat L_{2}\bar f^{(2)}_{\chi<}(t)\epsilon^2/2!+O(\epsilon^3).
\end{equation}
For (\ref{lde_chi}) holding in wider region, 
we require that the optimal values of $(p_{1}, p_{2}, t)$ are given by the simultaneous conditions,
\begin{eqnarray}
\hat L_{1}\hat L_{2}\bar f^{(1)}_{\chi<}&=&0,\label{first_d_chi1}\\
\hat L_{1}\hat L_{2}\bar f^{(2)}_{\chi<}&=&0,\label{first_d_chi2}\\
\hat L_{1}\hat L_{2}\bar f^{(3)}_{\chi<}&=&0.
\label{first_d_chi3}
\end{eqnarray}
We call this kind of conditions extended principle of minimum sensitivity (PMS) \cite{steve}.  
One should thus understand that the $5$th derivative $\bar f^{(5)}_{\chi<}$ participates the estimation task.  However, the counter part $\bar f^{(5)}_{\chi>}$ which shall be substituted in the place of $\bar f^{(5)}_{\chi<}$ does not show expected scaling to $25$th order (See FIG. 1(b)).  To circumvent the difficulty, we remind that frequently used value of $\theta\sim 0.5$ implies that $p_{3}=2(\theta/2\nu)=2p_{1}$ and $p_{3}$ is close to $p_{2}=1/2\nu$.  Thus, we may use the operator $\hat L_{1}\hat L_{3}$ rather than $\hat L_{1}\hat L_{2}$ and consider
\begin{equation}
\hat L_{1}\hat L_{3}\bar f_{\chi<}=\frac{\gamma}{2\nu}.
\label{lde_chi2}
\end{equation}
The advantage of this LDE is that the left-hand-side includes only two arguments $(p_{1}, t)$ and we need just
\begin{eqnarray}
\hat L_{1}\hat L_{3}\bar f^{(1)}_{\chi<}&=&0,\label{first_d_chi3a}\\
\hat L_{1}\hat L_{3}\bar f^{(2)}_{\chi<}&=&0,
\label{first_d_chi4}
\end{eqnarray}
where the highest derivative order is $4$th.  More explicitly, (\ref{first_d_chi3a}) and (\ref{first_d_chi4}) are written as
\begin{equation}
\Big[1+\frac{3}{2p_{1}}\frac{d}{d\log t}+\frac{1}{2p_{1}^2}\Big(\frac{d}{d\log t}\Big)^2\Big]\bar f_{\chi<}^{(k)}=\bar f_{\chi<}^{(k)}+\frac{3}{2p_{1}}\bar f_{\chi<}^{(k+1)}+\frac{1}{2p_{1}^2}\bar f_{\chi<}^{(k+2)}=0,\quad (k=1,2).
\end{equation}
By substituting $\bar f^{(k)}_{\chi>}$ into $\bar f^{(k)}_{\chi<}$ relying on the scaling behaviors captured in $\bar f^{(k)}_{\chi>}$ and solving simultaneous equations at the highest expansion order $25$th, we have two sets of solutions, $(p_{1}^*, t^*)=(1.42858, 0.11722)$ and $(2.49926, 0.11791)$ \cite{comment1}.  These sets respectively lead that
\begin{equation}
\omega=1.39999412\cdots,\quad 0.80023659\cdots.
\label{omega}
\end{equation}
The first solution in (\ref{omega}) is too large compared with the average $\omega\sim 0.84$ quoted in ref.\cite{peli}.  It shall be excluded after the successive estimation of $p_{2}=1/(2\nu)$ as we can see below.

Using two values of $\omega$ obtained in (\ref{omega}), the exponent $\nu$ is computed  with $f_{\beta}$ and associated LDE.  The derivatives of $\bar f_{\beta>}$ show clear expected scaling behaviors up to the third order.  At fourth order, we may regard that the scaling behavior has just began.  This means that when the unknown constant $\nu$ is involved, safely used LDE may be again $K=2$ case.  Then, as in the previous case, we have two options for the estimation of $\nu$.   They are expressed by
\begin{equation}
\hat L_{1} \hat L_{3} \bar f_{\beta<} = -\frac{1}{2\nu},\label{nulde1}
\end{equation}
or
\begin{equation}
\hat L_{1} \hat L_{2} \bar f_{\beta<} = -\frac{1}{2\nu}.
\label{nulde2}
\end{equation}
Consider first (\ref{nulde1}) which involves no adjustable parameter except $t$, the estimation point.   We simply search for the stationary point of $\hat L_{1} \hat L_{3}  \bar f_{\beta>}$ or otherwise the point of least variation ($\bar f_{\beta<}$ should be replaced by $\bar f_{\beta>}$) with the substitution of $\omega$ in (\ref{omega}).  At the highest order $24$th, this yields for respective $\omega$ values
\begin{equation}
\nu=0.63593,\quad 0.62928.
\label{nu1}
\end{equation}

Next we turn to (\ref{nulde2}).  With (\ref{omega}), $\nu$ estimation may be straightforwardly carried out by the use of extended PMS condition, $\hat L_{1} \hat L_{2} \bar f_{\beta}^{(1)} =0$ and $\hat L_{1} \hat L_{2} \bar f_{\beta}^{(2)}=0$ or in explicit terms,
\begin{equation}
\Big[1+(\frac{1}{p_{1}}+\frac{1}{p_{2}})\frac{d}{d\log t}+\frac{1}{p_{1}p_{2}}\Big(\frac{d}{d\log t}\Big)^2\Big]\bar f_{\beta<}^{(k)}=0,\quad (k=1,2).
\label{lde_self}
\end{equation}
This set involves $4$th order derivative and two unknown variables $p_{2}$ and $t$.  Rather than including $\bar f_{\beta}^{(4)}$, we have found more effective way of estimation by noting the structure of Wegner expansion:  Observing that the right-hand-side of (\ref{nulde2}) can be written as $-p_{2}$, the LDE (\ref{nulde2}) is expressed as
\begin{equation}
\Big[1+(\frac{1}{p_{1}}+\frac{1}{p_{2}})\frac{d}{d\log t}+\frac{1}{p_{1}p_{2}}\Big(\frac{d}{d\log t}\Big)^2\Big]\bar f_{\beta<}=-p_{2}.
\label{lde_self2}
\end{equation}
Hence, just adding the $k=1$ condition of (\ref{lde_self}), set up of obtaining $p_{2}$ and $t$ is completed.  Since the highest order of the derivatives is just the third one, this self-consistency recipe is expected to work better than the normal extended PMS condition.   Solving the simultaneous equations with the replacement of $\bar f_{\beta>}$ into $\bar f_{\beta<}$, we obtain at $24$th order that
\begin{equation}
\nu=0.63508,\quad 0.62948,
\label{nu2}
\end{equation}
for respective two $\omega$ values.  

The both results in (\ref{nu1}) and (\ref{nu2}) deduced from $\omega=1.39999$ are larger compared to the standard value.  More important point is that the set of values $(\omega, \nu)=(1.39999\cdots, 0.635\cdots)$ conflicts with the presumption that $p_{2}$ and $p_{3}$ are close with each others:   From the values of $\omega$ and $\nu$ obtained, we have $\theta=\omega\nu\sim 0.889$.  This is too large so breaks the $\omega$ estimation using  $L_{1}L_{3}$ instead of $L_{1}L_{2}$ (This becomes valid only when $\theta\sim 0.5$).  In contrast, the set $(0.80023659\cdots, 0.629\cdots)$ gives $\theta\sim 0.5036(1)$ and is consistent with the presumption.
  Thus, we keep only the solution $\omega=0.80023659\cdots$ and discard larger one henceforth.  We show in FIG.2 the plot of $\hat L_{1}\hat L_{2}\bar f_{\beta>}$ at $N=24$ where $\omega=0.80023$ and $\nu=0.62948$ are substituted.  Though not so wide, the plateau region is observed at which the value of $-p_{2}$ is estimated.
\begin{figure}
\centering
\includegraphics[scale=0.8]{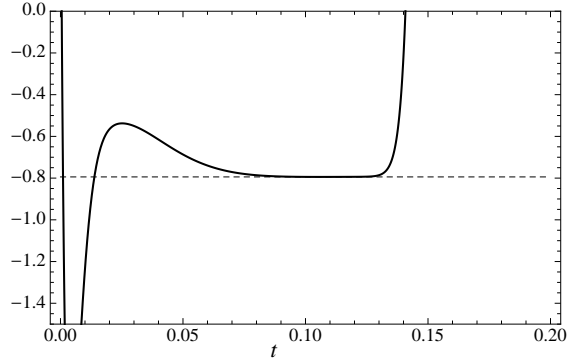}
\caption{Plot of $\hat L_{1}\hat L_{2}\bar f_{\beta>}$ at the highest order $24$th with $\omega=0.80023659$ and $\nu=0.62948475$ which is obtained from the point of view of self-consistency.  Dotted line represents the stationary value of $\hat L_{1}\hat L_{2}\bar f_{\beta>}$ which agrees with $-p_{2}^{*}=-0.7943004$.}
\end{figure}  
  
Turning to the proper $\nu$ estimation with $\omega=0.80023659\cdots$, we have presented two values in (\ref{nu1}) and (\ref{nu2}).  They were derived from the LDEs $\hat L_{1} \hat L_{3}  \bar f_{\beta}= -1/2\nu$ and $\hat L_{1} \hat L_{2}  \bar f_{\beta}= -1/2\nu$ respectively.   Now we have obtained that $2\theta>1$ and thus confirmed that $p_{2}<p_{3}$ within our results.  Though the two exponents are really close with each other and the two estimations of $\nu$ too, the results from $\hat L_{1} \hat L_{2}  \bar f_{\beta}=-1/(2\nu)$ would be the best one.  Thus, we place emphasis on the result,
\begin{equation}
\nu=0.62948.
\label{nu_final}
\end{equation}     

Having fixed most appropriate estimation of $\omega$ and $\nu$, we address to the estimation of $\eta$ and $\gamma$.  Proper LDE is $\hat L_{1}\hat L_{2} \bar f_{\chi}=\gamma/(2\nu)$.  To reduce the bias from $\nu$, we first treat $p_{2}=1/(2\nu)$ as an adjustable parameter.  We then solve $\hat L_{1}\hat L_{2} \bar f_{\chi>}^{(k)}=0$ for $k=1,2$ and obtain $(p_{2}^{*}, t^{*})=(0.8002365924\cdots, 0.11791552\cdots)$.  Then, we obtain $\gamma/(2\nu)=\hat L_{1}\hat L_{2} \bar f_{\chi>}|_{t=t^{*},p_{2}=p_{2}^{*}}=0.9816460\cdots$.  
From Fisher's relation $\gamma/2\nu=1-\eta/2$, we thus obtain 
$\eta=0.03671$.
Then, from the fixed $\nu$ in (\ref{nu_final}) and $\eta$ just obtained, we find
$
\gamma=1.23585
$.  
Another recipe of estimation is to bias $p_{2}$ by the value (\ref{nu_final}).  This recipe gives at the least variation point of $\hat L_{1}\hat L_{2} \bar f_{\chi} |_{p_{1}=p_{1}^{*},p_{2}=p_{2}^{*}}$, $\gamma/(2\nu)=0.98145537\cdots$.  
Then we obtain $\eta=0.03709$ and $\gamma=1.23561$ \cite{comment2}.   It is not clear to us which way of estimation is more reliable.  Fortunately, the two sets are close with each other, and we satisfy ourselves to take average of the two and conclude that
\begin{equation}
\eta=0.0369,\quad \gamma=1.2357.
\end{equation}

The value of $\eta$ is slightly larger compared to the standard one, $\eta=0.0364(5)$, quoted in ref.\cite{peli}.  As for $\gamma$, 
our value is slightly lower than that quoted in ref.\cite{peli} $\gamma=1.2372(5)$.  In our estimation, $\gamma$ is affected by $\omega$, $\nu$ and $\eta$.  Actually, if $\nu=0.6301$ quoted in ref.\cite{peli} is used in $\gamma/(2\nu)$, we obtain $\gamma=1.2371$ and $1.2368$ in the respective estimations.   In conclusion, our approach is basically working good.

\section{Biased estimation}
We have so far performed estimations of critical exponents in the self-contained and unbiased manner.  Here we present biased estimation and compare the results with those of world average at present.  
As an input we use $\omega$ quoted as the summary in ref.\cite{peli},
\begin{equation}
\omega=0.84(4).
\label{omega2}
\end{equation}
Then, we estimate $\nu$ under the self-consistent condition via (\ref{nulde2}).  The result at $N=24$ reads
\begin{equation}
\nu=0.63013\,(-0.00065, +0.00058),
\label{nu_biased}
\end{equation}
where the minus deviation implies result for $\omega=0.80$ and the plus deviation, $\omega=0.88$.  Note that the indicated range comes from the uncertainty of $\omega$ and not from some statistical origin.  To compare our result with the central value of accumulated results $\nu=0.6301(4)$ \cite{peli}, our estimate is in good agreement.

The exponent $\eta$ is estimated by using (\ref{first_d_chi1}) and (\ref{first_d_chi2}).  Then, $\gamma$ is obtained from the result and (\ref{nu_biased}).  At $N=25$, in the treatise of $p_{2}$ being adjustable, we obtain that
$\eta=0.03695\,(-0.00024, +0.00022)$ and $\gamma=1.23698\,(+0.00015, -0.00014)$.  In the case of substituting the values of $p_{1}$ and $p_{2}$ via $\omega=0.84(4)$ and $\nu=0.63013$, $\eta=0.03758\,(-0.00044,+0.00040)$ and $\gamma=1.23658\,(+0.00028, -0.00025)$.  Here, 
the first and the last numbers in the parenthesis show the deviation at $\omega=0.80$ and $0.88$, respectively.  The obtained values of $\gamma$ and $\eta$ show slight discrepancy with $\gamma=1.2372(5)$ and $\eta=0.0364(5)$ quoted in ref.\cite{peli}.  However, inclusion of the uncertainty in the parenthesis allows us to conclude that the estimation biased by $\omega$ is roughly consistent with the average of existing literatures quoted in ref.\cite{peli}.

\section{Concluding remarks}
We first comment on the trend of $\omega$ estimate as the order of expansion increases.  Nontrivial solution of $\omega$ from $\hat L_{1}\hat L_{3}\bar f_{\chi>}$ appears from $21$st order.  The results read $0.92800, 0.86046, 0.82024, 0.80411, 0.80023$ from $21$st to $25$th orders.  The point is that the sequence shows decreasing trend.   It might take place that the limit of the sequence would be slightly smaller than $0.8$.  If $\omega$ is below $0.79$, it would mean $\theta<0.5$ and lead $p_{1}<p_{3}<p_{2}$.   It is yet fair to say that any conclusion cannot be drawn since the trend would change to increasing, as  often encountered in examples presented in \cite{yam,yam3}.  

As the next remark, it is in order to mention on the use of $3$rd order LDE.  One might consider that $\hat L_{1}\hat L_{2}\hat L_{3}\bar f_{\beta}$ and $\hat L_{1}\hat L_{2}\hat L_{3}\bar f_{\chi}$ may be useful to estimate $1/2\nu$ and $\eta=2-\gamma/\nu$.  The first point is the estimation of $\omega$:   This $3$rdLDE includes three unknown variables $p_{1}$, $p_{2}$ and $t$ the estimation point.  Then, the needed highest derivative is $6$th and $25$th order series is too short to indicate scaling.  We therefore suffice ourselves with the value obtained with $2$ndLDE, the second value in  (\ref{omega}).   Then, naive recipe is to use proper values $\omega=0.80023659\cdots$ and $\nu=0.62948\cdots$ in $p_{k}$ $(k=1,2,3)$ and search for stationary or least variation points.   As for $\eta$, the plateau becomes narrower compared to the $2$-parameter ansatz, and the stationary value remains almost same as the $2$ndLDE result.   For $\nu$, $\hat L_{1}\hat L_{2}\hat L_{3}\bar f_{\beta}$ supplies no plateau and the scaling behavior is disappeared.  Next, if we use $p_{2}=1/2\nu$ as an adjustable parameter and adopt extended PMS for $\eta$ estimation, we obtain $p_{2}=\infty$, meaning $t^{-p_{2}}=0$.   In general, this takes place when the order of large mass expansion is not large enough.   Also in $\nu$ estimation, self-consistency requirement gives no solution.  Thus, $25$th order series is not enough for $3$rdLDE analysis.   Both of the previous and this issues would be settled by further higher order computation.

In the present paper, we have examined and demonstrated that our approach based upon the large mass expansion with crucial assist of $\delta$-expansion provided the results consistent with the accumulated results discussed in ref.\cite{peli}.  It would be worth of emphasizing that computations via the high temperature expansion can yield accurate results by itself.  To summarize the best results,
\begin{eqnarray}
\omega&=&0.8002,\\
\nu&=&0.6295,\\
\eta&=&0.0369,\\
\gamma&=&1.2357.
\end{eqnarray}
Estimation biased by $\omega$ has also given accurate results.  All results stemming from our approach are in good agreement with the standard values of critical exponents.


\begin{thebibliography}{00}
\bibitem{comment1}  As a reference, we have tried to search for $\omega$ by using $\hat L_{1}\hat L_{3} \bar f_{\beta>}^{(k)}=0$ $(k=1,2)$.  Only at $N=23$, we obtained solutions for $\omega$, $\omega=1.31085\cdots$ and $0.69469\cdots$.   Since other orders failed to give $\omega$ and the weak behavior of the scaling of $\bar f_{\beta>}$ at $23$rd order, we discarded these solutions.
\bibitem{comment2}It is possible to estimate $\eta$ from (\ref{lde_chi2}).  By the substitution of the value $\omega=0.80023659\cdots$ and we found a stationary point indicating $\gamma/(2\nu)$.  The result of $\eta$ deduced shows good agreement with the value.
\end{thebibliography}
\end{document}